\documentclass[aps,prl,twocolumn,superscriptaddress,showpacs]{revtex4}
\usepackage{epsfig}
\newcommand{\beq}{\begin{equation}} 
\newcommand{\eeq}{\end{equation}} 
 
\newcommand{\ve}{\vert} 
\newcommand{\ran}{\rangle} 
\newcommand{\lan}{\langle}

\begin{document}            

\title{Ising-like dynamics and frozen states in systems of ultrafine magnetic particles}
\author{Stefanie Russ} 
\affiliation{Institut f\"ur Theoretische Physik III, Justus-Liebig-Universit\"at
  Giessen, D-35392 Giessen, Germany}
\author{Armin Bunde} 
\affiliation{Institut f\"ur Theoretische Physik III, Justus-Liebig-Universit\"at
  Giessen, D-35392 Giessen, Germany}
\date{\today}

\draft
\begin{abstract} 
We use Monte-Carlo simulations to study aging phenomena and the occurence of spinglass phases in systems of single-domain ferromagnetic nanoparticles under the combined influence of dipolar interaction and anisotropy energy, for different combinations of positional and orientational disorder. We find that the magnetic moments oriente themselves preferably parallel to their anisotropy axes and changes of the total magnetization are solely achieved by $180$ degree flips of the magnetic moments, as in Ising systems. Since the dipolar interaction favorizes the formation of antiparallel chain-like structures, antiparallel chain-like patterns are frozen in at low temperatures, leading to aging phenomena characteristic for spin-glasses. Contrary to the intuition, these aging effects are more pronounced in ordered than in disordered structures.
\end{abstract}
\pacs{75.75.+a, 75.40.Mg, 75.50.Lk, 75.50.Tt
}
\maketitle

\section{Introduction}

In the last decade, systems of ultrafine magnetic nanoparticles have received considerable interest, due both to their important technological applications (mainly in magnetic storage and recordings) and their rich and often unusual experimental behavior, which is related to their role as a complex mesoscopic system \cite{battle02,kleemannferro04}. It has been discussed controversially in the past, under which circumstances these systems are able to show spin-glass phases. While experiments on disordered magnetic materials present indications of a spin-glass phase \cite{kleemannferro04,jonsson,Chantrell} or of a glassy-like random anisotropy system \cite{Luo}, the situation is less clear on the theoretical side. Simulations of the zero-field cooling (ZFC) and field-cooling susceptibility showed no indication of a spin-glass phase \cite{PortoPRL,porto05}. In contrast, simulations on aging \cite{Andersson} (on a simplified system, where the dipolar interaction was only considered up to a cut-off radius) and magnetic relaxation \cite{ulrich,russ06} favorize the spin-glass hypothesis, but the structure of the frozen history-dependent states as well as the actual mechanism leading to them has not yet been clarified.

In this letter, in order to clarify these questions, we use Monte Carlo simulations \cite{Nowak} to study aging phenomena on a large variety of systems of ultrafine magnetic nanoparticles (see Fig.~\ref{bi:geo}). Our simulations do not only point to the existence of frozen history-dependent states at low temperatures that are characteristic for spin glasses, but also yield an insight into the structure of the frozen states and the underlying dynamics. We find that under the combined influence of dipolar and anisotropy energy, the magnetic moments have a tendency to align in an Ising-like manner either parallel or antiparallel to their anisotropy axes and change their directions by $180$ degree flips as in Ising systems.  
This way, chain-like structures are formed where all magnetic moments point into the same direction and neighboring chains have the tendency to oriente themselves in an antiparallel way. These topological chains that freeze in at low temperatures, form simple straight lines, when the particles are arranged on the sites of a cubic lattice \cite{russ06} and form complex winded curves, when the arrangement of the particles is liquid-like. As a consequence, if a small external magnetic field is applied, the magnetic moments can follow the field more easily in a disordered system than in the ordered configuration. This leads, contrary to the intuition, to more pronounced aging effects (characteristic for spin glasses) in ordered than in disordered structures.

\unitlength 1.85mm
\vspace*{0mm}
\begin{figure}
\begin{picture}(40,38)
\put(2,20){\makebox{\includegraphics[width=3.cm]{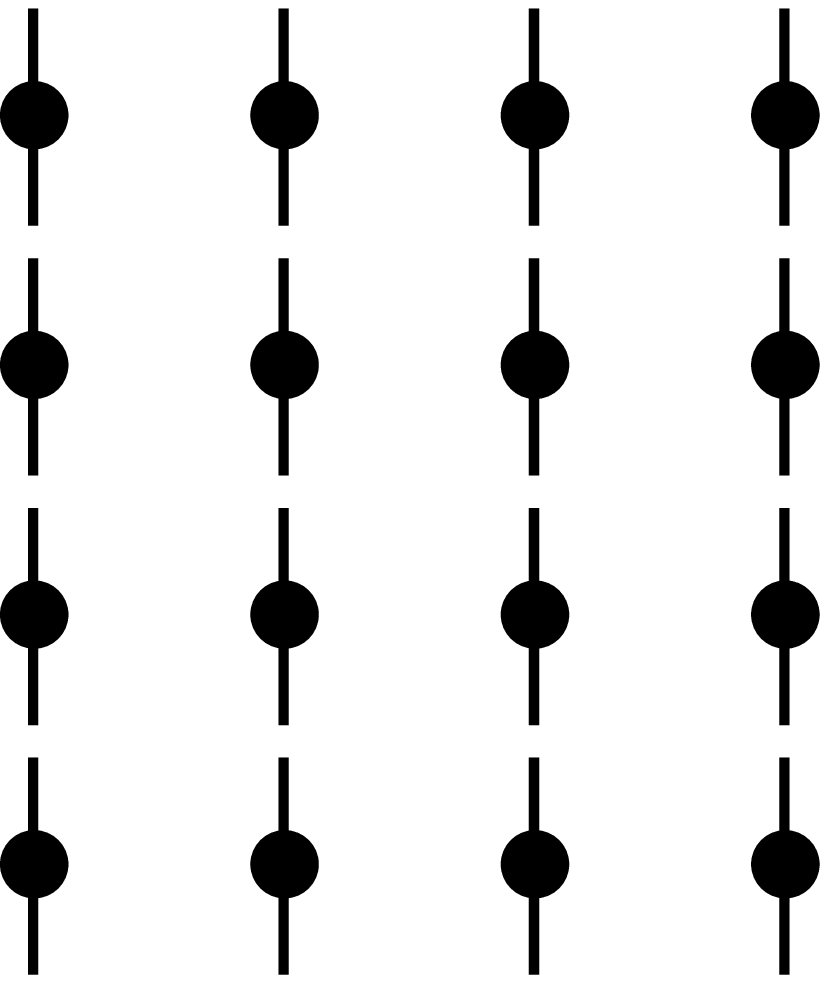}}}   
\put(2,0){\makebox{\includegraphics[width=3.cm]{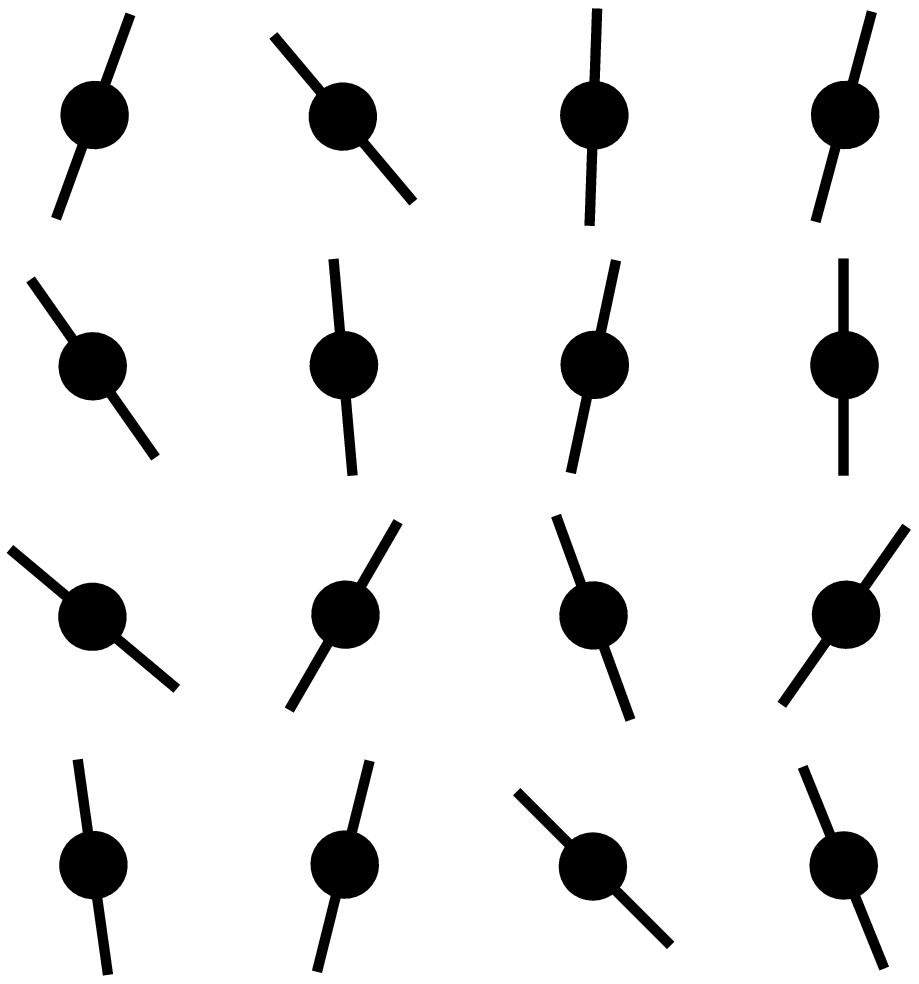}}}   

\put(25,20){\makebox{\includegraphics[width=3.cm]{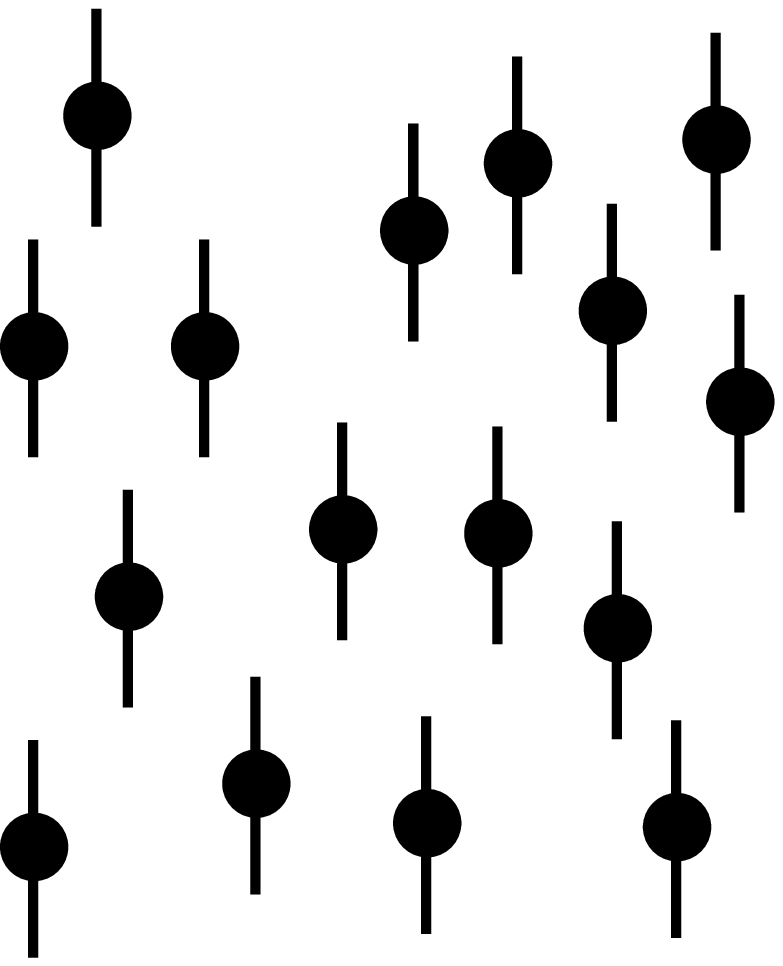}}}   
\put(25,0){\makebox{\includegraphics[width=3.cm]{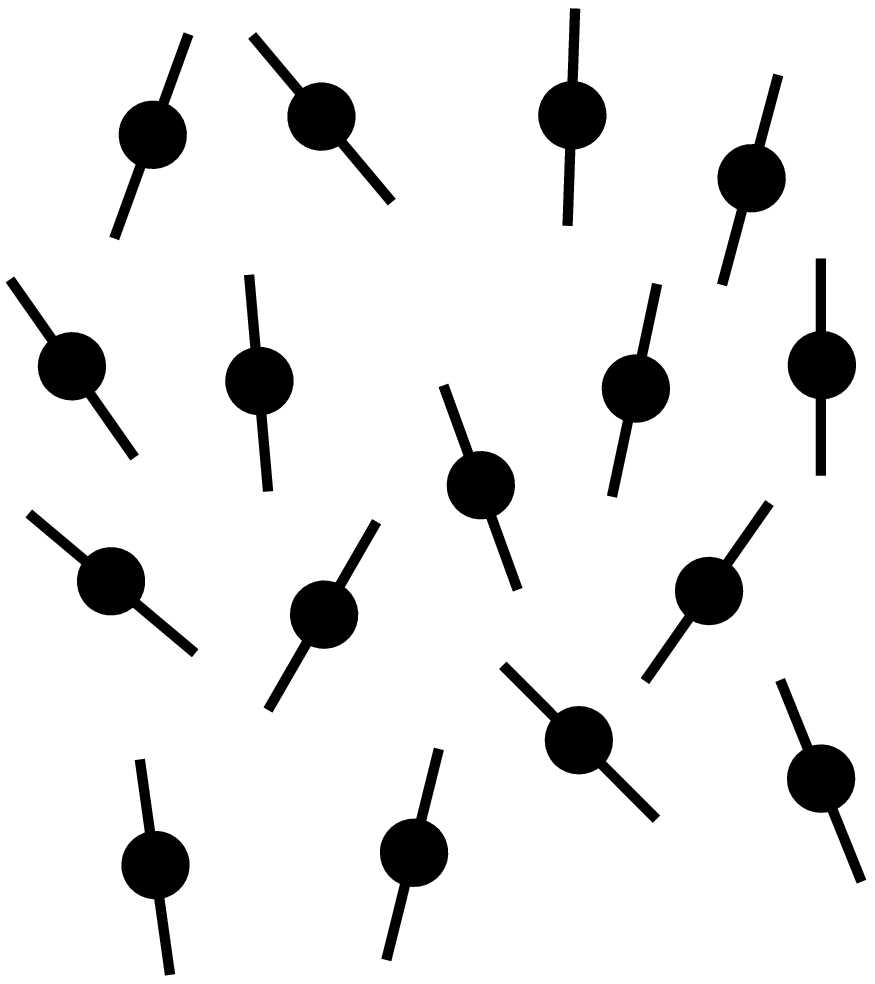}}}   

\put(-1,35){\makebox(1,1){\bf\Large (a)}} 
\put(22,35){\makebox(1,1){\bf\Large (b)}} 
\put(-1,15){\makebox(1,1){\bf\Large (c)}} 
\put(22,15){\makebox(1,1){\bf\Large (d)}} 

\end{picture}
\caption[]{\small Two-dimensional sketches of the geometries considered in this paper: (a) cubic arrangement of the particles and all anisotropy axes aligned into the $z$-direction, (b) liquid-like arrangement and all axes arranged, (c) cubic arrangement and all axes randomly oriented and (d) liquid-like arrangement and all axes randomly oriented. In the simulations, the systems were three-dimensional ($64$ particles per cube).
}
\label{bi:geo} 
\end{figure}

\section{Model System and Numerical Simulations}

For the numerical calculations, we focus on the same model as in earlier papers \cite{PortoPRL,ulrich}, which (i) assumes a coherent magnetization rotation within the anisotropic particles, and (ii) takes into account the magnetic dipolar interaction between them. Every particle $i$ of volume $V_i$ is considered to be a single magnetic domain $\vec\mu_i$ with all its atomic magnetic moments rotating coherently and the $V_i$ are taken from a Gaussian distribution of width $\sigma_V=0.4$ and $\lan V\ran=1$ (see also ~\cite{ulrich,PortoPRL}). This results in a constant absolute value $\ve\mu_i\ve=M_sV_i$ of the total magnetic moment of each particle, where $M_s$ is the saturation magnetization. The energy of each particle consists of three contributions: anisotropy energy, dipolar interaction and magnetic energy of an external field. We assume a temperature independent uniaxial anisotropy energy $E_A^{(i)}=-KV_i((\vec \mu_i \vec n_i)/\ve \vec{\mu_i} \ve)^2$, where $K$ is the anisotropy constant and the unit vector $\vec n_i$ denotes the easy directions.
Eventually, the magnetic moments are coupled to an external field $H$ leading to the additional field energy $E_H^{(i)}=- \vec \mu_i \vec H$. Finally, the energy of the magnetic dipolar interaction between two particles $i$ and $j$ separated by $\vec r_{ij}$ is given by $E_D^{(i,j)}=(\vec \mu_i \vec \mu_j)/r_{ij}^3 -3(\vec \mu_i \vec r_{ij})(\vec \mu_j\vec r_{ij})/r_{ij}^5$. Adding up the three energy contributions and summing over all $N$ particles we obtain the total energy
\beq
E=\sum_i^N E_A^{(i)} + \sum_i^N E_H^{(i)} +\frac{1}{2}\sum_i^N\sum_{j\ne i}^N E_D^{(i,j)}.
\eeq
In the Monte Carlo simulations we concentrate on samples of $N=L^{3}$ particles placed inside a cube of side length $L=4$ and average over $1000$ configurations. During the simulations, both, the positions of the particles and their easy axes are kept fixed.
The unitless concentration $c$ is defined as the ratio between the total volume $\sum_i V_i$ occupied by the particles and the volume $V_s$ of the sample. Here, we focus on the concentration $c/c_0\approx 0.3$, where $c_0=2K/M_s^2$ is a dimensionless material-dependent constant, $c_0\sim 1.4$ for iron nitride and $c_0\sim 2.1$ for maghemite nanoparticles \cite{ulrich}. We also tested systems with higher concentrations $c/c_0\approx 0.4$ and (the extremely high concentration) $c/c_0\approx 0.6$ and found that the results remain qualitatively unchanged. The temperature is measured in units of the reduced temperature $\tilde T \equiv 1/(2\beta KV)$, where $2KV$ is the height of the anisotropy barrier and $\beta=1/(k_BT)$. Similarly, the magnetic field is measured in units of the anisotropy field $H_a=2K/M_s$. The relaxation of the individual magnetic moments is simulated by the standard Metropolis algorithm \cite{metropolis}. In contrast to \cite{Andersson}, where dipole interactions between the particles were only considered up to a cut-off radius, we calculate the interaction energies by the Ewald sum method with periodic boundary conditions in $x$, $y$ and $z$-direction \cite{PortoPRL,ewald} and thus are able to account fully for the long-range character of the dipole forces. The magnetic moment $\vec \mu_i$ is characterized by the spherical angles $\theta_i$ and $\varphi_i$ relative to a coordinate frame, where the $z$-axis is parallel to the external field \cite{chamber,ngai,ulrich}. To study the magnetic relaxation we determine as a function of time $t$ (number of Monte Carlo steps) for each particle $i$ the angle $\theta_i$ between the magnetic moment $\vec{\mu_i}$ and the $z$-axis, from which we obtain the relevant quantities, as e.g. the normalized magnetization,
\beq\label{mag}
m(t) = \frac{1}{N}\sum_{i=1}^N \frac{V_i}{\langle V \rangle} \cos\theta_i(t).
\eeq

To obtain the orientation of $\vec\mu_i$ relative to $\vec n_i$, we introduce the ''orientational order parameter'' $O_\mu\equiv\lan\vert\vec\mu_i\vec n_i\vert\ran$, i.e. the average of the absolute values of the scalar product $\vec\mu_i\vec n_i$ over all $N$ particles and all configurations. $O_\mu$ does not distinguish between the parallel and the antiparallel alignment. It is equal to zero when all $\vec\mu_i$ are perpendicular to their axes $\vec n_i$ and equal to $1$ if they are all parallel or antiparallel to them. 

\unitlength 1.85mm
\vspace*{0mm}
\begin{figure}
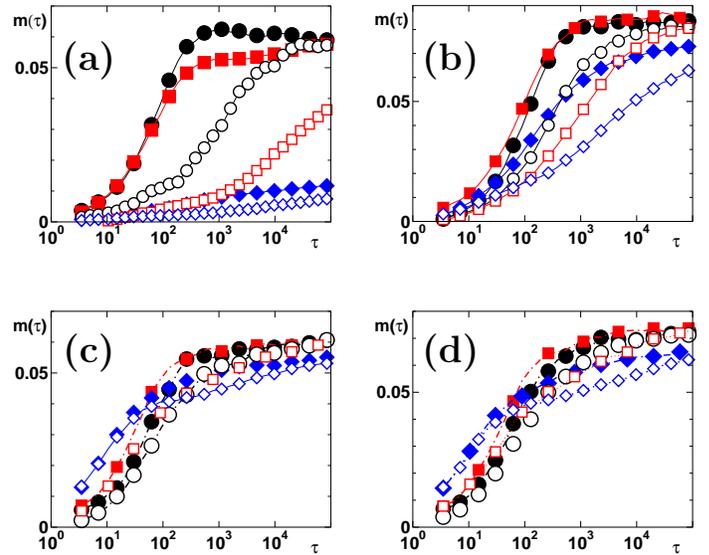

\begin{picture}(40,40)
\put(-4,22){\makebox{\includegraphics[width=4.3cm]{age.cub.ali.eps}}}   
\put(-4,0){\makebox{\includegraphics[width=4.3cm]{age.cub.axr.eps}}}   

\put(22,22){\makebox{\includegraphics[width=4.3cm]{age.dis.ali.eps}}}   
\put(22,0){\makebox{\includegraphics[width=4.3cm]{age.dis.axr.eps}}}   

\put(1,35){\makebox(1,1){\bf\Large (a)}} 
\put(27,35){\makebox(1,1){\bf\Large (b)}} 
\put(1,14){\makebox(1,1){\bf\Large (c)}} 
\put(27,14){\makebox(1,1){\bf\Large (d)}} 

\end{picture}
\caption[]{\small (Colors online) The magnetization $m(\tau)$ after waiting times $t_w=0$ (filled symbols) and $t_w=10000$ Monte Carlo steps (open symbols) is plotted versus $\tau$ (number of Monte Carlo steps with applied external field) for (a) cubic lattice and aligned axes, (b) liquid-like system and aligned axes, (c) cubic system and random axes and (d) liquid-like systems and random axes for the reduced temperatures $\tilde T=k_BT/(2KV)=5$ (black symbols, circles), $\tilde T=1/10$ (red symbols, squares) and $\tilde T=1/40$ (blue symbols, diamonds).
}
\label{bi:fig_age_c03} 
\end{figure}

To study aging phenomena, we determine the magnetization in a ZFC simulation. First, starting in a random configuration of the magnetic moments, the system is cooled down in the absence of an external field, from $T=\infty$ to a reduced temperature $\tilde T$ with a constant cooling rate of $\Delta \beta/\Delta t=0.1$, corresponding to $400$ Monte Carlo steps for $\tilde T=1/40$ and $10$ steps for $\tilde T=1$. Second, the cooling process is stopped at $\tilde T$ and the system is allowed to relax for a certain waiting time $t_w$. Finally, in the third step, a small external field $h=0.1 H_a$ is applied in $z$-direction. The magnetization $m(\tau)$ is determined as a function of $\tau\equiv t-t_w$ (number of Monte Carlo steps after switching on the field). Aging effects are represented by differences between the $m(\tau)$-curves for different $t_w$ and occur, when many different relaxation rates exist in the system, so that after a given waiting time $t_w$, the system has only partly relaxed towards equilibrium. 
Experimentally, aging effects have already been found in several spin-glasses, as e.g. in Permalloy/alumina granular films \cite{vincent}, rare-earth manganates \cite{kundu05}, $Cu Mn$ spin-glasses \cite{lundgren83}, multilayer systems \cite{kleemannferro05} and in $Fe_3N$ nanoparticle systems \cite{sasaki}.

\section{Numerical Results}

Figure~\ref{bi:fig_age_c03} shows $m(\tau)$ for the systems of Fig.~\ref{bi:geo} without waiting time, $t_w=0$ (filled symbols), and for $t_w=10^4$ (open symbols). The different colors (and symbols) stand for three different temperatures $\tilde T=1/5$, $1/10$ and $1/40$. Clearly, all curves show aging effects, similar to the experimental results of Refs.~\cite{vincent,lundgren83,kundu05}. Systems with no or only small $t_w$ follow the external field faster than the systems with longer waiting times, indicating that the longer relaxation leads to more stable chains.
The aging effects are most pronounced for those systems where all anisotropy axes are oriented into the direction of the external field (Fig.~\ref{bi:fig_age_c03}(a,b)) and less pronounced but still visible for the systems with disordered anisotropy axes (Fig.~\ref{bi:fig_age_c03}(c,d)). In these systems with orientational disorder, the $m(\tau)$ curves coincide for small $\tau$ and show aging effects only after a certain crossover time (close to $10^{2}$ Monte Carlo steps). This indicates that in these systems a certain fraction of dipoles does not belong to quasi-stable chain-like structures and can follow the external field nearly instantaneously, independentely of the waiting time and thus dominate the short-time behavior. The aging effects decrease with increasing $\tilde T$, when the order is destroyed by the thermal fluctuations. 

\unitlength 1.85mm
\vspace*{0mm}
\begin{figure}
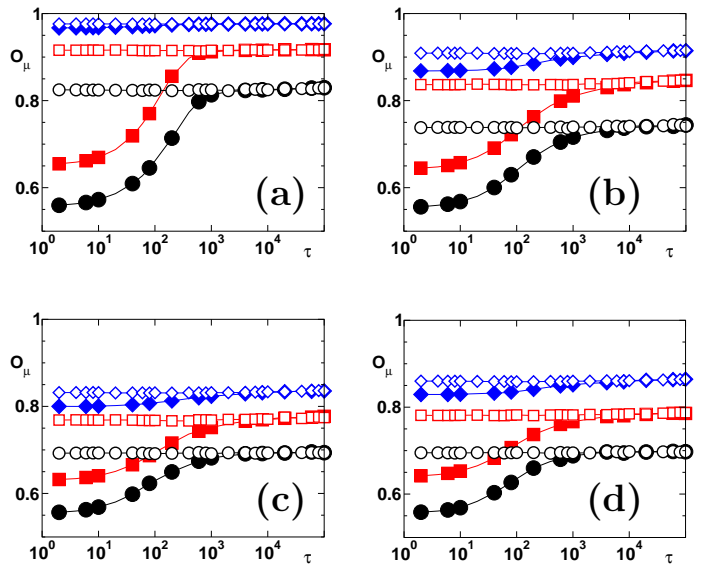

\begin{picture}(40,40)
\put(-4,22){\makebox{\includegraphics[width=4.3cm]{order.cub.ali.eps}}}   
\put(-4,0){\makebox{\includegraphics[width=4.3cm]{order.cub.axr.eps}}}   

\put(22,22){\makebox{\includegraphics[width=4.3cm]{order.dis.ali.eps}}}   
\put(22,0){\makebox{\includegraphics[width=4.3cm]{order.dis.axr.eps}}}   

\put(15,26){\makebox(1,1){\bf\Large (a)}} 
\put(39,26){\makebox(1,1){\bf\Large (b)}} 
\put(15,4){\makebox(1,1){\bf\Large (c)}} 
\put(39,4){\makebox(1,1){\bf\Large (d)}} 

\end{picture}
\caption[]{\small (Colors online) The order parameter $O_\mu$ after a waiting time $t_w=0$ (filled symbols) and $t_w=10000$ (open symbols) is plotted versus $\tau$ (number of Monte Carlo steps) for the same geometries, temperatures, system parameters and symbols and colors as in Fig.~\ref{bi:fig_age_c03}.
}
\label{bi:fig_ageorder} 
\end{figure}

\unitlength 1.85mm
\vspace*{0mm}
\begin{figure}
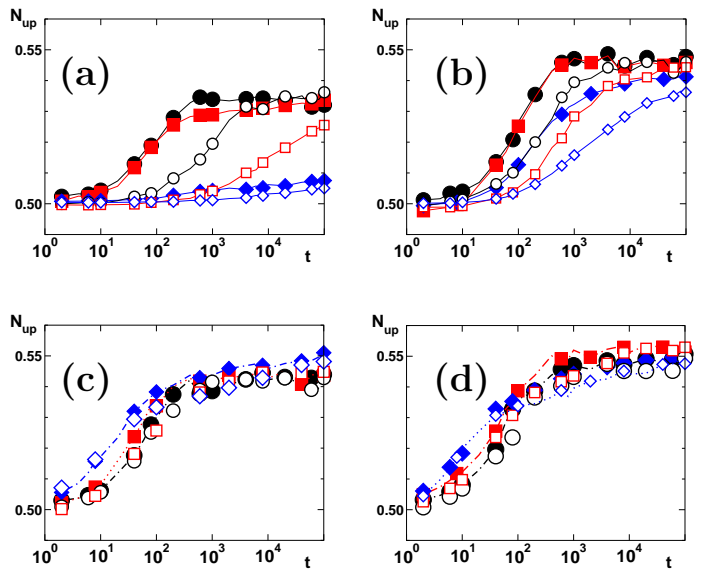

\begin{picture}(40,40)
\put(-4,22){\makebox{\includegraphics[width=4.3cm]{numberz.cub.ali.eps}}}   
\put(-4,0){\makebox{\includegraphics[width=4.3cm]{numberz.cub.axr.eps}}}   

\put(22,22){\makebox{\includegraphics[width=4.3cm]{numberz.dis.ali.eps}}}   
\put(22,0){\makebox{\includegraphics[width=4.3cm]{numberz.dis.axr.eps}}}   

\put(1,35){\makebox(1,1){\bf\Large (a)}} 
\put(28,35){\makebox(1,1){\bf\Large (b)}} 
\put(1,13){\makebox(1,1){\bf\Large (c)}} 
\put(28,13){\makebox(1,1){\bf\Large (d)}} 

\end{picture}
\caption[]{\small (Colors online) The percentage $N_{\rm{up}}$ of particles per system pointing upwards after waiting times $t_w=0$ (filled symbols) and $t_w=10000$ (open symbols) is plotted versus $\tau$ (number of Monte Carlo steps) for the same geometries, temperatures, system parameters and symbols and colors as in Fig.~\ref{bi:fig_age_c03}.
}
\label{bi:number} 
\end{figure}

In order to understand the dynamical behavior in a more microscopic way, we compare $m(\tau)$ with the time-dependence of the corresponding orientational order parameters $O_\mu(\tau)$. Figure~\ref{bi:fig_ageorder} shows $O_\mu$ in the 3rd step of the aging process for $t_w=0$ and $t_w=10000$ (filled and open symbols, respectively) and for the same geometries as before (see Fig.~\ref{bi:geo}). The figure shows that quite contrary to the expectation, apart from a slight minimum at intermediate $\tau$, $O_\mu$ is constant in time for the systems of $t_w=10000$. Without waiting time, the curves start at much smaller values of $O_\mu$, but increase rapidly until they reach at a crossover time $\tau_c$ of about $10^{3}$ Monte Carlo steps the common plateau value. In the plateau regime, the dipolar moments $\vec\mu_i$ are either oriented parallel or antiparallel to their easy axes $\vec n_i$ and do therefore flip only between these two directions. Accordingly, the value of $O_\mu$ does neither depend on the external field nor on the functional form of $m(\tau)$. Since $O_\mu(\tau)$ stays constant for large $t_w$ or $\tau>\tau_c$, while $m(\tau)$ increases with time (see Fig.~\ref{bi:fig_age_c03}), the $\vec\mu_i$ have already reached their parallel or antiparallel position and can only perform spin flips by $180$ degrees, thereby increasing $m(\tau)$ and leaving $O_\mu$ unchanged. To make this point still clearer, we plot in Fig.~\ref{bi:number} the percentage $N_{\rm{up}}$ of particles pointing upwards, i.e. with $\vartheta_i<\pi/2$, again for the geometries of Fig.~\ref{bi:geo}. The similarity between Fig.~\ref{bi:number} and Fig.~\ref{bi:fig_age_c03} is obvious, showing that the number of the magnetic moments oriented upwards determine the shape of $m(\tau)$.

\unitlength 1.85mm
\vspace*{0mm}
\begin{figure}
\begin{picture}(40,25)
\put(2,0){\makebox{\includegraphics[width=6cm]{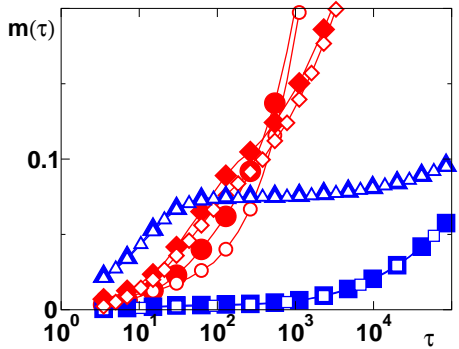}}}   
\end{picture}
\caption[]{\small (Colors online) The magnetization $m(\tau)$ after a waiting time $t_w=0$ (filled symbols) and $t_w=10000$ (open symbols) for $\tilde T=1/10$ (red symbols) and $\tilde T=1/40$ (blue symbols) of systems with aligned and randomly oriented anisotropy axes (red circles and diamonds, respectively for $\tilde T=1/10$ and blue squares and triangles respectively for $\tilde T=1/40$) are plotted versus $\tau$ (number of Monte Carlo steps) for systems without dipole-interaction. 
}
\label{bi:fig_agenoww} 
\end{figure} 

We therefore arrive at a remarkably simple Ising-like dynamics of these ultrafine magnetic particles. The amount of aging is directly related to the degree of order a system can achieve during $t_w$. In the fully ordered system of Figs.~\ref{bi:geo}-\ref{bi:fig_ageorder}(a), after a long waiting time $t_w$, the $\vec\mu_i$ prefer to be aligned in stable chains \cite{russ06} along the $z$-direction and thus cannot follow an external field easily. Single magnetic moments inside a chain will hardly flip to the other side and flips of whole chains possess extremely large relaxation times. Without waiting time, on the other hand, the $\vec\mu_i$ are in unstable positions which allows them to follow the external field quite rapidly, leading to large aging effects in ordered systems. As Figs.~\ref{bi:fig_age_c03}(b-d) show, the situation is different in systems with positional and/or orientational disorder. The relaxation times for spin flips decrease with the amount of disorder, in particular with the amount of orientational disorder. When the chains are winded and aligned into different directions, they are less stable and possess a large variety of intermediate positions to flip to the other side. Accordingly, aging effects become weaker with increasing disorder. 

For illustration, we visualize the aging process in Fig.~\ref{bi:pattern} for the system with the highest order and the strongest aging effects, i.e. for the cubic system with aligned anisotropy axes.  For this visualization, we follow the definition of the transversal order parameter of Ref.~\cite{russ06}: each of the $L^2$ sites in the $xy$ plane can be either a $+$ site or a $-$ site, if a chain has already been formed and all magnetic moments in the chain point into the positive or negative $z$ direction, respectively (white sites). If this is not the case, the site is a $0$ site (grey sites). The figure shows that chains are quite obviously formed in the second step of the aging process during the waiting time $t_w$, as can most easily be seen by comparing Fig.~\ref{bi:pattern}(a), where $t_w=10000$ with \ref{bi:pattern}(d) where $t_w=0$. In (a), many chains are formed during $t_w$ that appear to be quite stable in the following 3rd step of the aging process (Fig.~\ref{bi:pattern}(b,c)), when an external field is applied in the $+$ direction. We can see that most of the $-$ chains persist in spite of the external field. The situation is different in Fig.~\ref{bi:pattern}(d-f), where only few chains exist at the end of the 2nd step of the aging process (Fig.~\ref{bi:pattern}(d)). Here, after switching on the external magnetic field, new chains can be built from the $0$ sites and the system therefore follows the field much easier than in Fig.~\ref{bi:pattern}(a-c).

Recently, it has been argued that also a broad distribution of anisotropy energy barriers might lead to aging effects in superparamagnetic systems \cite{sasaki}.
To show that these kinds of aging effects are in fact negligible compared with systems where both energy contributions are present, we have studied systems without dipole interaction (solely anisotropy energy) at temperatures $\tilde T=1/10$ and $1/40$. In this case, the particle positions play no role, so that the geometry of Fig.~\ref{bi:geo}(a) and (b) as well as (c) and (d) are physically identical. The results of $m(\tau)$ for these two geometries are shown in Fig.~\ref{bi:fig_agenoww} for the same aging procedure as before. The figure shows that the differences between the curves for $t_w=0$ and $t_w=10000$ are orders of magnitude smaller than in the systems with dipolar interaction. It is interesting to note that also for systems with only dipolar interaction, some kind of aging can be seen, but orders of magnitude smaller than for systems with both energy contributions.

\unitlength 1.85mm
\vspace*{0mm}
\begin{figure}
\begin{picture}(40,30)
\put(0,17){\makebox{\includegraphics[width=2.cm]{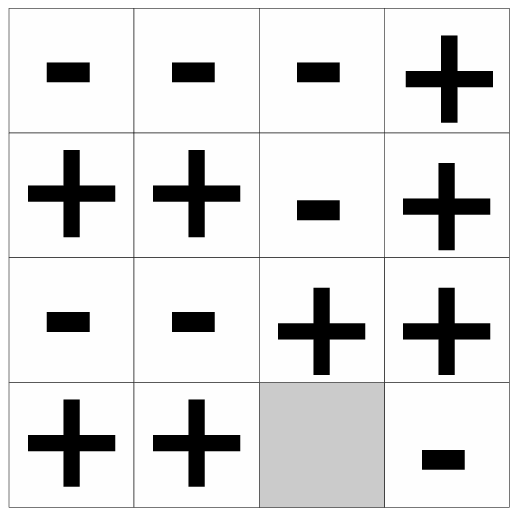}}}   
\put(13,17){\makebox{\includegraphics[width=2.cm]{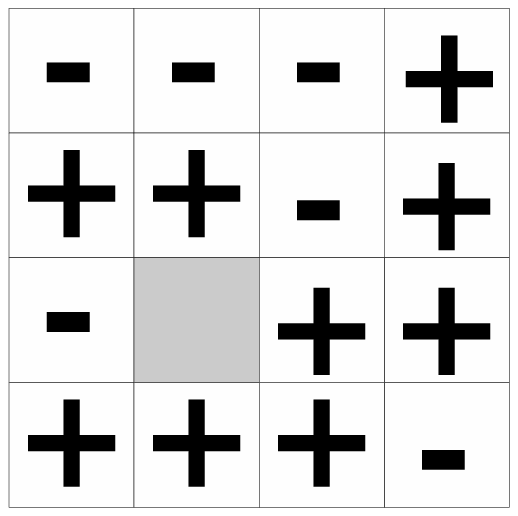}}}   
\put(26,17){\makebox{\includegraphics[width=2.cm]{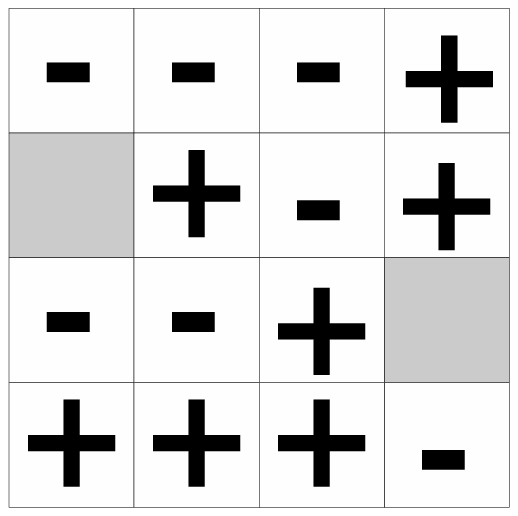}}}   

\put(0,0){\makebox{\includegraphics[width=2.cm]{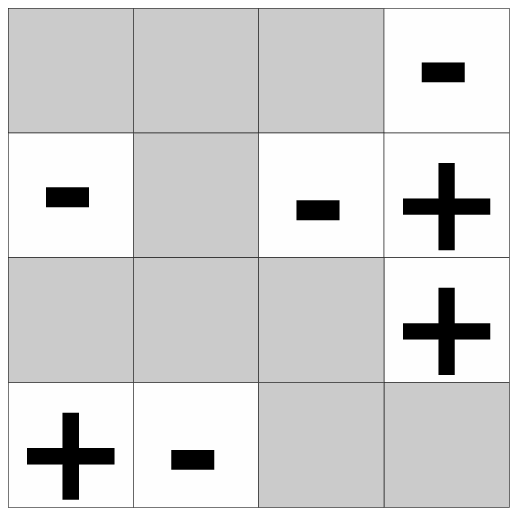}}}   
\put(13,0){\makebox{\includegraphics[width=2.cm]{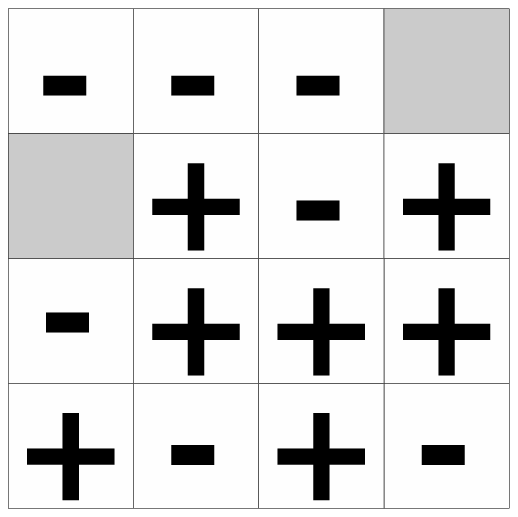}}}   
\put(26,0){\makebox{\includegraphics[width=2.cm]{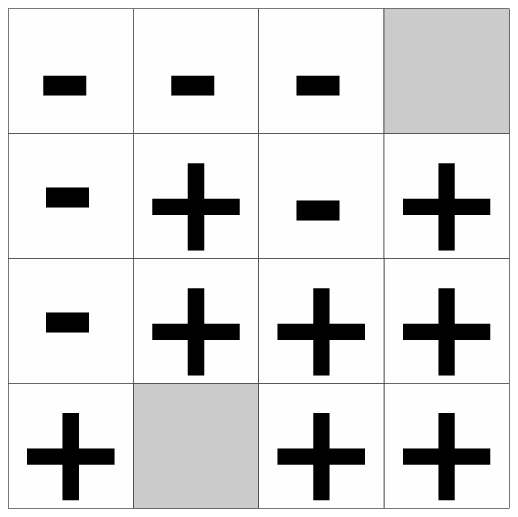}}}   

\put(1,30){\makebox(1,1){\bf\Large (a)}} 
\put(14,30){\makebox(1,1){\bf\Large (b)}} 
\put(28,30){\makebox(1,1){\bf\Large (c)}} 
\put(1,13){\makebox(1,1){\bf\Large (d)}} 
\put(14,13){\makebox(1,1){\bf\Large (e)}} 
\put(28,13){\makebox(1,1){\bf\Large (f)}} 

\end{picture}
\caption[]{\small Visualization of the chains perpendicular to the $xy$-plane in the cubic system with aligned anisotropy axes at $\tilde T=1/5$ for one typical system. The complete chains are indicated by white sites and by $+$ or $-$ signs, depending on the direction of the chain. Sites, where chains have not (yet) been built are indicated by the grey shade.
(a-c) System with waiting time $t_w=10000$, i.e. (a) after the cooling process and $t_w=10000$ (b,c) after an external magnetic field in the $+$ direction has been applied for (b) $1000$ and (c) $10000$ Monte Carlo steps.
(d-f) System without waiting time ($t_w=0$), i.e. (d) after the cooling process (and $t_w=0$) (e,f) after an external magnetic field in the $+$ direction has been applied for (e) $1000$ and (f) $10000$ Monte Carlo steps.
}
\label{bi:pattern} 
\end{figure}

In summary, analyzing the microscopic dynamics of ultrafine magnetic particles, we found that irrespective of the strength of the dipolar interaction, the dipoles oriente themselves either parallel or antiparallel to their anisotropy axes. We therefore arrive at a remarakably simple picture of the dipole dynamics, where similar to the Ising model, the $\vec\mu_i$ perform ''spin flips'' between these two orientations. Aging effects occur when after a certain waiting time, the magnetic dipoles have arranged themselves in stable configurations and flips of single magnetic moments are suppressed. These aging effects increase in a counter-intuitive way with the order of the system and are thus most pronounced in completely orderded systems with cubic arrangement of the particles and axes aligned into the direction of the magnetic field. 

\section{acknowledgements}

We gratefully acknowledge very valuable discussions with W. Kleemann and financial support from the Deutsche Forschungsgemeinschaft.


\begin{thebibliography} {50}
\bibitem{battle02} X. Batlle and A. Labarta, J. Phys. D 35, R15 (2002). 
\bibitem{kleemannferro04} Xi Chen, S. Sahoo, W. Kleemann, S. Cardoso and P.~P.~Freitas, Phys. Rev. B {\bf 70}, 172411 (2004).
\bibitem{jonsson} T. Jonsson, J. Mattsson, C. Djurberg, F. A. Khan, P. Nordblad, and P. Svedlindh, Phys. Rev. Lett. 75, 4138 (1995).
\bibitem{Chantrell} R. W. Chantrell, M. El-Hilo, and K. O Grady, IEEE Trans. Magn. {\bf 27}, 3570 (1991).
\bibitem{Luo} W. Luo, S. R. Nagel, T. F. Rosenbaum, and R. E. Rosensweig, Phys. Rev. Lett. {\bf 67}, 2721 (1991). 
\bibitem{PortoPRL} J. Garc´{i}a-Otero, M. Porto, J. Rivas, and A. Bunde, Phys. Rev. Lett. {\bf 84}, 167 (2000).
\bibitem{porto05} M. Porto, Eur. Phys. J. B {\bf 45}, 369 (2005).
\bibitem{Andersson} J.-O. Andersson et al., Phys. Rev. B {\bf 56}, 13983 (1997).
\bibitem{ulrich} M. Ulrich, J. Garc´{i}a-Otero, J. Rivas, and A. Bunde; Phys. Rev. B {\bf 67}, 024416 (2003).
\bibitem{russ06} S. Russ, A. Bunde, Phys. Rev. B {\bf 74}, 064426 (2006).
\bibitem{Nowak} U. Nowak, R. W. Chantrell, and E. C. Kennedy, Phys. Rev. Lett. {\bf 84}, 163 (2000).
\bibitem{metropolis} In every step, we select a particle $i$ at random and generate an attempted orientation of its magnetization, chosen in a spherical
segment around the present orientation with an aperture angle
$d\theta$ (see also Ref.~\cite{PortoPRL}). By varying $d\theta$, i.e. the maximum jump angle, it is possible to modify the rate of acceptance and to optimize the simulation.
As a compromise between simulations at low and high temperatures, we chose $d\theta=0.1$ for all simulations, independent of temperature, which refers to an accecptance rate between $0.5$ and $0.8$ for $\tilde T$ between $1/40$ and $1/5$. We also tested larger values of $d\theta$ with considerably lower acceptation rates and found that they did not change the final states significantly. 
\bibitem{ewald} M. P. Allen and D. J. Tildesley, {\it Computer Simulation of Liquids} (Clarendon, Oxford, 1987).
\bibitem{chamber} R. V. Chamberlin, G. Mozurkewich, and R. Orbach, Phys. Rev. Lett. 52, 867 (1984).
\bibitem{ngai} K. L. Ngai and U. Strom, Phys. Rev. B {\bf 38}, 10350 (1988). 
\bibitem{vincent} E. Vincent, Y. Yuan, J. Hamman, H. Hurdequint and F. Guevara; J. of Mag. and Mag. Mat. {\bf 161}209 (1996).
\bibitem{kundu05} A. K. Kundu, P. Nordblad and C. N. R. Rao; Phys. Rev. B {\bf 72}, 144423 (2005).
\bibitem{lundgren83} L. Lundgren, P. Svendlindh, P. Nordblad and O. Beckmann; Phys. Rev. Lett. {\bf 51}, 811 (1983).
\bibitem{kleemannferro05} S.~Bedanta, O.~Petracic, E.~Kentzinger, W.~Kleemann, U.~R\"ucker, A.~Paul, Th.~Br\"uckel, S.~Cardoso and P.~P.~Freitas, Phys. Rev. B {\bf 72}, 024419 (2005).
\bibitem{sasaki} M. Sasaki, P.E. J\"onsson, H. Takayama and H. Mamiya; Phys. Rev. B {\bf 71}, 104405 (2005). 
\end{thebibliography}
\end{document}